\documentclass[pre,showpacs,preprint]{revtex4}
\usepackage{bm}% bold math
\usepackage{amsmath}
\usepackage{graphicx}
\begin{document}

\title{Scaling and aging in the homogeneous cooling state of a granular fluid of hard particles}
\author{J.J. Brey, A. Prados, M.I. Garc\'{\i}a de Soria, and P. Maynar}
\affiliation{F\'{\i}sica Te\'{o}rica, Universidad de Sevilla,
Apartado de Correos 1065, E-41080, Sevilla, Spain}
\date{\today }

\begin{abstract}
The presence of the aging phenomenon in the homogeneous cooling
state (HCS)  of a granular fluid composed of inelastic hard spheres
or disks is investigated. As a consequence of the scaling property
of the $N$-particle distribution function, it is obtained that the
decay of the normalized two-time correlation functions slows down as
the time elapsed since the beginning of the measurement increases.
This result is confirmed by molecular dynamics simulations for the
particular case of the total energy of the system. The agreement is
also quantitative in the low density limit, for which an explicit
analytical form of the time correlation function has been derived.
Moreover, the reported results provide support for the existence of
the HCS as a solution of the N-particle Liouville equation.

\end{abstract}


\maketitle

\section{Introduction}
\label{s1} In many far from equilibrium states, it has been
observed that the relaxation (or response) rate decreases as the
``age'' of the process increases. Here ``age'' refers to the time
elapsed since the beginning of the considered experiment. Then, it
is said that the system ages, or that it exhibits an aging
phenomenon. A revision of the concept of aging in spin glasses and
in other systems can be found in ref. \cite{BCKyM97}.

One of the typical experiments for the analysis of aging in a given
physical system is the study of the two-time correlation functions
of some of its properties, since they  are related with the response
of the system to a given perturbation. Let $C_{AB}(t_{w},t)$ denote
the two-time correlation function of the magnitudes $A$ and $B$ of
the system, the former being measured at time $t_{w}$ and the latter
at time $t \geq t_{w}$. In a system at equilibrium,
$C_{AB}(t_{w},t)$ only depends on the the time difference $\tau=
t-t_{w}$, due to the time translational invariance. On the other
hand, in systems presenting aging, it depends on both $\tau$ and
$t_{w}$. It is important to stress that the aging phenomenon is more
than just the loss of the time translational invariance; it consists
in the relaxation of $C_{AB}(t_{w},t_{w}+\tau)$ slowing down as the
``waiting time'' $t_{w}$ increases.

The simplest aging phenomenon occurs when the two-time correlation
function depends only on the time ratio $\tau/t_{w}$, and it is
sometimes called ``full aging''. This behavior is exhibited by some
simple models, such as mean field model of spin glasses
\cite{B92,CKyR94} and the one-dimensional Ising model at zero
temperature \cite{PyB97,GyL00}. More complicated dependencies like
$\ln t / \ln t_{w}$ have also been found \cite{FyH88} and, with more
generality, behaviors of the form $h(t)/h(t_{w})$, with different
functions $h$ \cite{BCKyM97,Br94}.

Granular media are inherently non-equilibrium systems, due to the
dissipative character of the interactions between grains. There is a
continuous loss of kinetic energy and the system tends to a rest,
unless energy is being continuously injected into the system, for
instance through a vibrating wall. A kind of typical experiments
carried out in dense granular systems are those designed to
investigate compaction \cite{Ketal95}. Usually, the system is
submitted to a series of separated pulses or taps of a given short
duration. After each tap, the system is allowed to relax freely
until it reaches a metastable configuration with all the particles
at rest. In this state, there are many permanent contacts between
particles. Next, the system is tapped again and the process is
repeated many times. Properties of interest like the volume or the
energy are measured at each rest configuration. In this way, the
evolution of these properties as a function of time, measured in
number of taps, is obtained. For large number of taps, the system
tends to a steady state with a density that is a monotonically
decreasing function of the tapping intensity \cite{Ketal95,Netal98}.
The relaxation of the system towards the steady state is very slow
and clearly non-exponential. Furthermore, when the intensity of
tapping is changed cyclically, hysteresis phenomena in the density
are observed \cite{Netal98}. These behaviors are similar to those
found in structural glasses when submitted to cyclical variations of
their temperature \cite{Sch86} and are a quite strong evidence of
the presence of aging phenomena in compact granular media. Actually,
aging has been observed both experimentally \cite{JTMyJ00} and also
identified in simple models of compaction \cite{NyC99,ByP01}.

A completely different regime of granular systems are the so-called
granular gases, in which there are not permanent contacts between
particles, but they move freely and independently between
collisions. The simplest possible state of a granular gas is the
homogeneous cooling state (HCS), whose temperature decays
monotonically in time. At a microscopic level, the HCS is assumed to
be characterized by a phase space probability distribution in which
all the time dependence occurs through the temperature. For the case
of hard particles, this implies a scaling property and the
possibility of identifying some features of the time dependence of
many relevant properties of the system, without carrying out
explicit calculations. Among these properties are ensemble averages
as well as time-correlation functions \cite{DByL02,BGMyR04}. The
above peculiarities render the HCS a good candidate to investigate
in detail and by means of analytical methods the possible existence
of aging and, in the case of a positive answer, its origin and
properties. In spite of the above, it has not been until very
recently that attention has been devoted to this particular aspect
of the HCS \cite{AyP07}. The aim of this paper is twofold. Firstly,
to investigate in detail the possible existence of the aging
phenomenon in the HCS of a granular fluid of inelastic hard
particles, deepening into its origin. This is done both analytically
and by means of particle simulation methods. The theoretical
analysis will be based on the existence of the HCS at the level of
the $N$-particle distribution function and its scaling property, for
arbitrary density and inelasticity. Then, the accuracy of the
predictions following from the theory provides strong support for
the existence of the HCS in the context of a many body theory. This
is precisely the second aim of the paper.

The presentation here proceeds as follows. In Sec. \ref{s2}, some
general properties of the $N$-particle distribution function
defining the assumed HCS of a system of inelastic hard spheres or
disks are shortly reviewed. By means of an appropriate scaling of
the dynamics of the system,  general features of the time dependence
of the time-correlation functions can be identified. This is
discussed in Sec.\ \ref{s3}, and the results presented hold, in
principle, for a general pair of dynamical variables and arbitrary
density and inelasticity. It is shown that the system exhibits full
aging as a direct consequence of the scaling property of the HCS.
The particular case of the total energy of the system is considered
in Sec.\ \ref{s4}. The time self-correlation function of this
property in the HCS is known in detail quite accurately  in the low
density limit, then allowing to get very detailed information about
its time behavior. Comparison of the theoretical predictions with
molecular dynamics simulation results is also presented. A quite
good agreement is observed. Finally, Sec.\ \ref{s5} contains some
general comments and final remarks.

\section{The Homogeneous Cooling State of a granular fluid}
\label{s2} Consider a system on $N$ inelastic hard spheres ($d=3$)
or disks ($d=2$) of mass $m$ and diameter $\sigma$. The position and
velocity of particle $r$ at time $t$ will be denoted by ${\bm
q}_{r}(t)$ and ${\bm v}_{r}(t)$, respectively. The dynamics of the
system consists of free streaming, i.e. straight line motion along
the direction of the velocity until a pair of particles, $r$ and
$s$, is at contact, at which time their velocities ${\bm v}_{r}$,
${\bm v}_{s}$ change instantaneously to ${\bm v}_{r}'$, ${\bm
v}_{s}'$ according to
\begin{equation}
\label{2.1} {\bm v}^{\prime}_{r}= {\bm v}_{r} - \frac{1+\alpha}{2}
\left( \widehat{\bm \sigma} \cdot {\bm v}_{rs} \right) \widehat{\bm
\sigma},
\end{equation}
\begin{equation}
\label{2.2} {\bm v}^{\prime}_{s}= {\bm v}_{s} + \frac{1+\alpha}{2}
\left( \widehat{\bm \sigma} \cdot {\bm v}_{rs} \right) \widehat{\bm
\sigma},
\end{equation}
where ${\bm v}_{rs} \equiv {\bm v}_{r}-{\bm v}_{s}$ is the
relative velocity and $\widehat{\bm \sigma}$ is a unit vector
along ${\bm q}_{rs} \equiv {\bm q}_{r}- {\bm q}_{s}$ at contact.
Finally, $\alpha$ is the coefficient of normal restitution,
defined in the range $ 0 < \alpha \leq 1$ and that will be
considered here as a velocity-independent constant. The sequence
of free streaming and binary collisions determines a unique
trajectory of the system.

A macroscopic state is specified, at the statistical mechanics
level, in terms of a probability density $\rho (\Gamma, t)$, with
$\Gamma$ denoting a point in the $2Nd$ dimensional phase space of
the system, $\Gamma \equiv \left\{ {\bm q}_{1},{\bm v}_{1}, \ldots,
{\bm q}_{N}, {\bm v}_{N} \right\} $. The macroscopic variables of
interest are the average of microscopic observables $A (\Gamma)$ at
a given time $t$, defined in the two equivalent forms
\begin{equation}
\label{2.3}  \langle A(t) \rangle \equiv \int d \Gamma\, \rho
(\Gamma) e^{tL} A(\Gamma) = \int d \Gamma\, A(\Gamma) e^{-t
\overline{L}} \rho (\Gamma).
\end{equation}
In the above expressions, $L$ is the generator of the dynamics for
phase functions, while $\overline{L}$ is the generator of the
dynamics for distribution functions. Their expressions are
\begin{equation}
\label{2.4} L(\Gamma) \equiv \sum_{r=1}^{N} {\bm v}_{r} \cdot
\frac{\partial}{\partial \bm q}_{r}  + \frac{1}{2} \sum_{r=1}^{N}
\sum_{s \neq r}^{N} T(r,s),
\end{equation}
\begin{equation}
\label{2.5} \overline{L}(\Gamma) \equiv \sum_{r=1}^{N} {\bm v}_{r}
\cdot \frac{\partial}{\partial \bm q}_{r}  - \frac{1}{2}
\sum_{r=1}^{N} \sum_{s \neq r}^{N} \overline{T}(r,s),
\end{equation}
with the binary collisions operators $T(r,s)$ and
$\overline{T}(r,s)$ defined by
\begin{equation}
\label{2.6} T(r,s) \equiv \sigma^{d-1} \int d \widehat{\bm \sigma}\,
\Theta ( - \widehat{\bm \sigma} \cdot {\bm v}_{rs}) | \widehat{\bm
\sigma} \cdot {\bm v}_{rs}| \delta ({\bm q}_{rs}-{\bm \sigma})
\left( b_{rs}-1 \right),
\end{equation}
\begin{equation}
\label{2.7} \overline{T}(r,s) \equiv \sigma^{d-1} \int d
\widehat{\bm \sigma}\, \Theta ( \widehat{\bm \sigma} \cdot {\bm
v}_{rs}) | \widehat{\bm \sigma} \cdot {\bm v}_{rs}| \left[
\alpha^{-2} \delta ( {\bm q}_{rs} - {\bm \sigma}) b_{rs}^{-1}-
\delta ( {\bm q}_{rs} + {\bm \sigma}) \right].
\end{equation}
In these expressions, $d \widehat{\bm \sigma}$ is the solid angle
element corresponding to $\widehat{\bm \sigma}$, ${\bm \sigma}
\equiv \sigma \widehat{\bm \sigma} $, ${\bm q}_{rs} \equiv {\bm
q}_{r}-{\bm q}_{s}$, and $\Theta (x)$ is the Heaviside step
function. Moreover, $b_{rs}$ is the substitution operator that
replaces the velocities ${\bm v}_{r}$ and ${\bm v}_{s}$ to its right
by their ``postcollisional'' values accordingly with Eqs.\
(\ref{2.1}) and (\ref{2.2}). Thus for an arbitrary function $F$,
\begin{equation}
\label{2.8} b_{rs} F( {\bm v}_{r}, {\bm v}_{s}) = F ({\bm
v}^{\prime}_{r},{\bm v}^{\prime}_{s}).
\end{equation}
Finally, the operator $b_{rs}^{-1}$ is the inverse of $b_{rs}$, i.e.
it changes the velocities ${\bm v}_{r}$, ${\bm v}_{s}$ by their
``precollisional'' values,
\begin{equation}
\label{2.9} b_{rs}^{-1}  F( {\bm v}_{r}, {\bm v}_{s}) = F ({\bm
v}^{\prime \prime} _{r},{\bm v}^{\prime \prime}_{s}),
\end{equation}
\begin{equation}
\label{2.10} {\bm v}^{\prime \prime}_{r} ={\bm
v}_{r}-\frac{1+\alpha}{2 \alpha} \left( \widehat{\bm \sigma} \cdot
{\bm v}_{rs} \right) \widehat{\bm \sigma},
\end{equation}
\begin{equation}
\label{2.11} {\bm v}^{\prime \prime}_{s} ={\bm
v}_{r}+\frac{1+\alpha}{2 \alpha} \left( \widehat{\bm \sigma} \cdot
{\bm v}_{rs} \right) \widehat{\bm \sigma}.
\end{equation}

In summary, the dynamics of the probability distribution function in
phase space is governed by the Liouville equation
\begin{equation}
\label{2.12} \left( \frac{\partial}{\partial t} + \overline{L}
\right) \rho (\Gamma, t)=0.
\end{equation}
Due to the energy dissipation in collisions, there is no stationary
solution to the above Liouville equation, except in the elastic
limit $\alpha = 1$. A granular temperature $T$ is usually defined
from the average of the energy density. For a homogeneous state of a
system composed of hard particles,  it is given by
\begin{equation}
\label{2.13} T(t)=\frac{2}{Nd} \langle  E(t) \rangle,
\end{equation}
with $E$ being the total (kinetic) energy of the system. By using
Eq.\ (\ref{2.3}), it is found
\begin{equation}
\label{2.14} \frac{\partial T(t)}{\partial t} = - \zeta (t) T(t),
\end{equation}
where the ``cooling rate'' $\zeta(t)$ is identified as
\begin{equation}
\label{2.15} \zeta(t) = - \frac{2}{T(t) N d}\,  \langle LE(t)
\rangle \geq 0.
\end{equation}
Of course, there is a large class of time-dependent homogeneous
states, depending on the initial preparation. Here, it will be
assumed that, after a few collisions per particle, there is a
relaxation of the velocity distribution towards a ``universal''
form, characterized because its entire time dependence occurs
through the cooling temperature. This special state is called the
homogeneous cooling state (HCS) and, at the macroscopic level, it
is defined by a uniform number density $n_{h}$, a uniform but
time-dependent temperature $T_{h}(t)$, and a vanishing flow
velocity. Because of the absence of any additional microscopic
energy scale for hard particles, its distribution function has the
form
\begin{equation}
\label{2.16} \rho_{h}(\Gamma,t)= \left[ \ell v_{0}(t) \right]^{-Nd}
\rho_{h}^{*} \left( \left\{  \frac{{\bm q}_{rs}}{\ell}, \frac{{\bm
v}_{r}}{v_{0}(t)}; r,s=1, \ldots, N \right\} \right),
\end{equation}
where $v_{0}(t) \equiv \left( 2T_{h}/m \right)^{1/2}$ is a thermal
velocity and $\ell \equiv (n_{h} \sigma^{d-1} )^{-1}$ a
characteristic length proportional to the mean free path. The
above special form of the $N$-particle distribution function
allows to determine the temperature (and time) dependence of many
average properties without explicit calculations. This fact will
be actually exploited in the following.

The existence of the HCS solution to the Liouville equation has
already been assumed several times in the literature
\cite{BDyS97,GyvN00}. Although there is no direct proof of it, nor
a constructive solution of the Liouville equation for this state
has been developed, molecular dynamics (MD) simulations have shown
that some of its implications, e.g. the scaling law for the
temperature mentioned below, are observed in detail. Additional
support has been provided by means of a time  scale change that
transforms the assumed HCS distribution into a time-independent
distribution \cite{Lu01,BRyM04}. MD simulations seem to confirm
the existence of the steady state, that is reached after a few
collisions per particle. A more demanding evidence of the
existence of the HCS with a distribution function having the
scaling property given in Eq.\ (\ref{2.16}) is provided by the
results to be reported in this paper.

The temperature dependence of the cooling rate for the HCS,
$\zeta_{h}(t)$, can be determined by dimensional analysis to be
$\zeta_{h}[n_{h},T_{h}(t)] \propto T_{h}(t)^{1/2}$. Now, Eq.
(\ref{2.14}), particularized for the HCS of a system of hard
spheres or disks, can be integrated, to obtain the time dependence
of the temperature
\begin{equation}
\label{2.17} T_{h}(t) = T_{h}(t^{\prime}) \left[ 1 + \frac{
\zeta^{*} v_{0}(t') (t-t')}{2 \ell} \right]^{-2},
\end{equation}
with
\begin{equation}
\label{2.18} \zeta^{*} \equiv \frac{\ell \zeta_{h}(t)}{v_{0}(t)}
\end{equation}
being a dimensionless time-independent cooling rate. This
algebraic decay of the temperature of the HCS is known as the Haff
law \cite{Ha83}.

For the analysis of the HCS, it is useful to introduce the
dimensionless time scale $s$ defined through
\begin{equation}
\label{2.19} s(t)= \int_{0}^{t} dt'\, \frac{v_{0}(t')}{\ell}.
\end{equation}
Therefore, $s$ is proportional to the accumulated average number of
collisions per particle in the time interval $(0,t)$. In terms of
this new time variable, the cooling law (\ref{2.17}) becomes
\begin{equation}
\label{2.20} T(s) = T(s') e^{-(s-s') \zeta^{*}}.
\end{equation}
The $t$ and $s$ time scales are related through
\begin{equation}
\label{2.21} s=\frac{2}{\zeta^{*}} \ln \left[ 1+
\frac{\zeta_{h}(0)}{2}\ t \right],
\end{equation}
as can be directly seen by comparison of Eqs.\ (\ref{2.17}) and
(\ref{2.20}) or, equivalently, by direct integration of Eq.\
(\ref{2.19}).

\section{Time-correlation functions in the HCS}
\label{s3} As indicated in the previous section, the scaling
property of the distribution function of the HCS implies that the
time-dependence of many macroscopic properties of the system can be
identified without carrying out explicit calculations. Let
$A(\Gamma)$ be a homogeneous function of degree $a$ of the
velocities of the particles. Then, it is
\begin{eqnarray}
\label{3.1} A(\Gamma) \equiv A \left( \left\{ {\bm q}_{r},{\bm
v}_{r}; r=1,\ldots,N \right\} \right) & \equiv & A \left( \left\{
\ell {\bm q}_{r}^{*}, v_{0}(t) {\bm v}_{r}^{*}; r=1,\ldots,N
\right\} \right) \nonumber \\
&= & v_{0}^{a}(t) A  \left( \left\{ \ell {\bm q}_{r}^{*}, {\bm
v}_{r}^{*}; r=1,\ldots,N \right\} \right),
\end{eqnarray}
where ${\bm q}^{*}_{r} \equiv {\bm q}_{r} /\ell$ and ${\bm
v}^{*}_{r} \equiv {\bm v}_{r}/v_{0}(t)$. Examples of this kind of
properties are the center of mass velocity or the total energy of
the system. The average value of $A$ in the HCS is
\begin{equation}
\label{3.2}  \langle A(t) \rangle_{h}= \int d\Gamma\, A(\Gamma)
\rho_{h}(\Gamma,t) =v_{0}^{a}(t) \langle A \rangle^{*}_{h},
\end{equation}
where
\begin{equation}
\label{3.3} \langle A \rangle^{*}_{h} \equiv \int d \Gamma^{*}
\rho^{*}_{h}(\Gamma^{*}) A \left( \left\{ \ell {\bm q}_{r}^{*},{\bm
v}_{r}^{*}; r=1, \ldots, N \right\} \right),
\end{equation}
and $\Gamma^{*} \equiv \left\{ {\bm q}^{*}_{r}, {\bm v}^{*}_{r};
r=1, \ldots, N  \right\}$. Thus all the time dependence of $
\langle A(t) \rangle_{h}$ is in the factor $v_{0}^{a}(t)$.

Suppose next that $B(\Gamma)$ is also a homogeneous function of the
velocities of degree $b$,
\begin{equation}
\label{3.4} B \left( \Gamma \right) = v_{0}^{b}(t) B  \left(
\left\{ \ell {\bm q}_{r}^{*}, {\bm v}_{r}^{*}; r=1,\ldots,N
\right\} \right),
\end{equation}
and consider the HCS time-correlation function for $A$ and $B$
defined as
\begin{equation}
\label{3.4a} C_{AB}(t,t') \equiv \langle A(t) B(t') \rangle_{h} -
\langle A(t) \rangle_{h} \langle B(t') \rangle_{h},
\end{equation}
for $t \geq t' \geq 0$. By carrying out the transformation to
dimensionless variables, it can be shown that \cite{DByL02}
\begin{equation}
\label{3.5} \langle A(t) B(t') \rangle_{h} = v_{0}^{a}(t)
v_{0}^{b}(t') \langle A(s-s') B \rangle^{*}_{h},
\end{equation}
with
\begin{equation}
\label{3.6} \langle A(s) B \rangle^{*}_{h} = \int d \Gamma^{*}\,
\rho_{h}^{*}(\Gamma^{*})  A \left( \left\{ \ell {\bm q}_{r}^{*},
{\bm v}_{r}^{*}\right\},s \right) B \left( \left\{ \ell {\bm
q}_{r}^{*}, {\bm v}_{r}^{*} \right\} \right).
\end{equation}
Here
\begin{equation}
\label{3.7} A \left( \left\{ \ell {\bm q}_{r}^{*}, {\bm
v}_{r}^{*}\right\},s \right) = e^{s \mathcal{L}^{*}} A \left(
\left\{ \ell {\bm q}_{r}^{*}, {\bm v}_{r}^{*}\right\}\right),
\end{equation}
where $\mathcal{L}^{*}$ is the new generator for the dynamics of the
phase functions,
\begin{equation}
\label{3.8} \mathcal{L}^{*} (\Gamma^{*}) \equiv \frac{\zeta^{*}}{2}
\sum_{r=1}^{N} {\bm v}^{*}_{r} \cdot \frac{\partial}{\partial {\bm
v}_{r}^{*}}  + L^{*} (\Gamma^{*}) ,
\end{equation}
\begin{equation}
\label{3.9} L^{*}(\Gamma^{*}) = \frac{\ell}{v_{0}(t)}  L  (\Gamma)=
[L(\Gamma)]_{\left\{ {\bm q}_{r}={\bm q}_{r}^{*}, {\bm v}_{r}={\bm
v}_{r}^{*} \right\}, \sigma=\sigma^{*} }\, .
\end{equation}
The first term on the right hand side of Eq.\ (\ref{3.8}), is due to
the time-dependent scaling of the velocities with $v_{0}(t)$. Use of
Eqs.\ (\ref{3.2}) and (\ref{3.5}) into Eq.\ (\ref{3.4a}) yields
\begin{equation}
\label{3.10} C_{AB}(t,t') = v_{0}^{a}(t) v_{0}^{b}(t') C_{AB}^{*}
(s-s'),
\end{equation}
\begin{equation}
\label{3.11} C_{AB}^{*}(s) = \langle A(s) B \rangle_{h}^{*} -
\langle A \rangle_{h}^{*} \langle B \rangle _{h}^{*}\, .
\end{equation}
To identify the aging phenomena clearer, it is convenient to
normalize the correlation function to unity for $t=t'$, by defining
a relaxation function $\phi_{AB}(t,t')$ as
\begin{eqnarray}
\label{3.12} \phi_{AB} (t,t') & \equiv &
\frac{C_{AB}(t,t')}{C_{AB}(t',t')} = \left[
\frac{v_{0}(t)}{v_{0}(t')} \right]^{a} \phi_{AB}^{*}
(s-s'; \zeta^{*}) \nonumber \\
& = & e^{- (s-s')\, \frac{a \zeta^{*}}{2}} \phi_{AB}^{*} (s-s';
\zeta^{*}).
\end{eqnarray}
Upon writing the last equality above, use has been made of Eq.\
(\ref{2.20}). The dimensionless relaxation function $\phi_{AB}^{*}$
above is
\begin{equation}
\label{3.13} \phi_{AB}^{*} (s; \zeta^{*}) = \frac{ \langle A(s)B
\rangle^{*}_{h} -\langle A \rangle_{h}^{*} \langle B
\rangle_{h}^{*}}{ \langle A B \rangle_{h}^{*} - \langle A
\rangle_{h}^{*} \langle B \rangle_{h}^{*}}.
\end{equation}
It follows from Eq. (\ref{3.12}) that the two-time correlation
function depends on time only through the difference $s-s'$.
Consequently, there is no aging when time is measured in the
dimensionless scale $s$. In ref. \cite{AyP07}, the aging property
of the velocity time-autocorrelation function of a granular gas of
inelastic hard particles was investigated by means of MD
simulations. Time was measured by the average  cumulated number of
collisions per particles that, as said above, is proportional to
the dimensionless time scale $s$ used here. The simulations
indicate that the velocity autocorrelation function,
$C_{vv}(s,s^{\prime})$ in the language used here, depends both on
$s^{\prime}$ and $s-s^{\prime}$, and from this feature the authors
conclude that the system exhibits aging. Of course, the analysis
developed here applies for the case of the velocity
time-autocorrelation function, corresponding to $a=b=1$ and,
therefore, aging should not be expected according to the results
derived above.  This apparent discrepancy seems to occur because
of a wrong use of the physical concept of aging in ref.
\cite{AyP07}. As pointed out in the Introduction, for the
existence of aging, it is not enough the dependence of the
time-correlation function on both $s^{\prime}$ and $s-s^{\prime}$.
In fact, the analysis developed here leads to
\begin{equation}
\label{3.14} C_{AB}(s,s') = C_{AB}(s',s') \phi_{AB}(s,s') =f(s')
e^{-\frac{a \zeta^{*}}{2} (s-s')} \phi_{AB}^{*} (s-s'; \zeta^{*}).
\end{equation}
Nevertheless, all the dependence on $s'$ occurs in the prefactor
$f(s')=C_{AB}(s^{\prime},s^{\prime})$ and, therefore, there is no
real aging, since the decaying rate is always the same and only
the initial value changes with $s'$ for constant $s-s'$.

On the other hand, the aging phenomenon shows up in the original
time scale $t$ in the limits $\zeta_{h}(0)t' \gg 1$ and
$\zeta_{h}(0) t \gg 1$. In this regime, it follows from Eq.
(\ref{2.21}) that
\begin{equation}
\label{3.15} s-s' \sim \frac{2}{\zeta^{*}} \ln \frac{t}{t`},
\end{equation}
and, therefore, Eq.\ (\ref{3.12}) takes the form
\begin{equation}
\label{3.16} \phi_{AB}(t,t') = \left( \frac{t'}{t} \right)^{a}
\phi_{AB}^{*} \left[ \frac{2}{\zeta^{*}} \ln \frac{t'}{t};
\zeta^{*} \right] \equiv  F \left(\frac{t}{t'}; \zeta^{*} \right),
\end{equation}
valid for $t,t' \gg \zeta_{h}(0)^{-1}$. This result is the
mathematical expression of the aging phenomenon. The normalized
time-correlation function depends on the initial and final times,
$t'$ and $t$, only through their quotient $t/t'$, so the system
exhibits full aging, as defined in the Introduction. The remaining
parameter determining the long time behavior of the correlation
function is the dimensionless cooling rate $\zeta^{*}$. It is worth
to stress the generality of this result. No limitation on the degree
of inelasticity or density has been introduced. In fact, both
magnitudes are relevant in determining, through the value of
$\zeta(0)$,  the time region in which the aging behavior predicted
by Eq. (\ref{3.14}) is to be expected. The only hypothesis made here
is the existence of the HCS with a distribution function having the
scaling form given in Eq. (\ref{2.16}). Of course, the system is
assumed to stay in that state for all the relaxation time $t$
considered. This requires that the HCS, in addition to exist, be
stable, at least in a determined region of parameters. The extensive
measurements of the velocity correlation function by means of MD
simulations reported in ref. \cite{LByD02} show that it is possible
to cover a wide range of density and inelasticity in which the
observed homogenous state appears to be stable. This is confirmed by
the simulation results to be presented here in the next section.

\section{Energy time correlation function in the HCS for a dilute
granular gas}

\label{s4} Theoretical predictions for the explicit form of the
function $\phi^{*}_{AB}(s,\zeta^{*})$ are scarce in the
literature. An exception is the total energy autocorrelation
function, $C_{EE}(t,t')$, for a dilute granular gas of hard
particles. By projecting the Liouville equation onto the
hydrodynamic modes in the low density limit, an analytical
expression for $C_{EE}$ was derived in ref. \cite{BGMyR04},
\begin{equation}
\label{4.1} C_{EE}(t,t') = N T(t) T(t') e (\alpha)
e^{-(s-s^{\prime}) \frac{\zeta^{*}}{2}},
\end{equation}
where $e(\alpha)$ is a given function of only the coefficient of
restitution. Then,
\begin{equation}
\label{4.2} \phi_{EE}(t,t') \equiv
\frac{C_{EE}(t,t')}{C_{EE}(t',t')} = \frac{T(t)}{T(t')} e^{-
(s-s') \frac{\zeta^{*}}{2}}.
\end{equation}
Since the energy $E$ is a homogenous function of degree $a=2$ of the
velocities of the particles, comparison of the above expression with
the general result given in Eq.\ (\ref{3.12}) leads to the
identification
\begin{equation}
\label{4.3} \phi_{EE}^{*}(s;\zeta^{*}) = e^{-\frac{\zeta^{*}}{2}\,
s}.
\end{equation}
Therefore, using Eq.\ (\ref{3.16}), it follows that the aging
behavior of the total energy of a dilute granular gas of hard
particles is given by
\begin{equation}
\label{4.4} \phi_{EE} (t,t') = e^{-\frac{3 \zeta^{*}}{2}\,
(s-s^{\prime})} \sim \left( \frac{t'}{t} \right)^{3},
\end{equation}
for $t,t' \gg \zeta_{h}(0)^{-1}$. An explicit expression for the
cooling rate of a dilute granular gas has been derived from the
Boltzmann equation in the so-called first Sonine approximation
\cite{GyS95}. This expression has been shown to accurately agree
with the numerical results obtained by means of the direct
simulation method of the Boltzmann equation \cite{BRyC96,BRyM04} in
the thermal velocity region, i.e. for velocities of the order of
$v_{0}(t)$.

Due to the simplicity of $\phi^{*}_{EE}$ in the present case, the
asymptotic form of the normalized time correlation function for the
energy, $\phi_{EE}(t,t')$, does not depend on the value of
$\zeta^{*}$, what implies that it is independent of the density
$n_{h}$ and of the coefficient of restitution $\alpha$.
Consequently, if $\phi_{EE}(t,t')$ is plotted as a function of
$t/t'$ for times in the range $t,t' \gg \zeta(0)^{-1}$, the curves
corresponding to different densities and inelasticities  should tend
to collapse on a unique one. Of course, the density range to which
this result applies is restricted by the validity of the Boltzmann
description.

\begin{figure}
\includegraphics[scale=0.5]{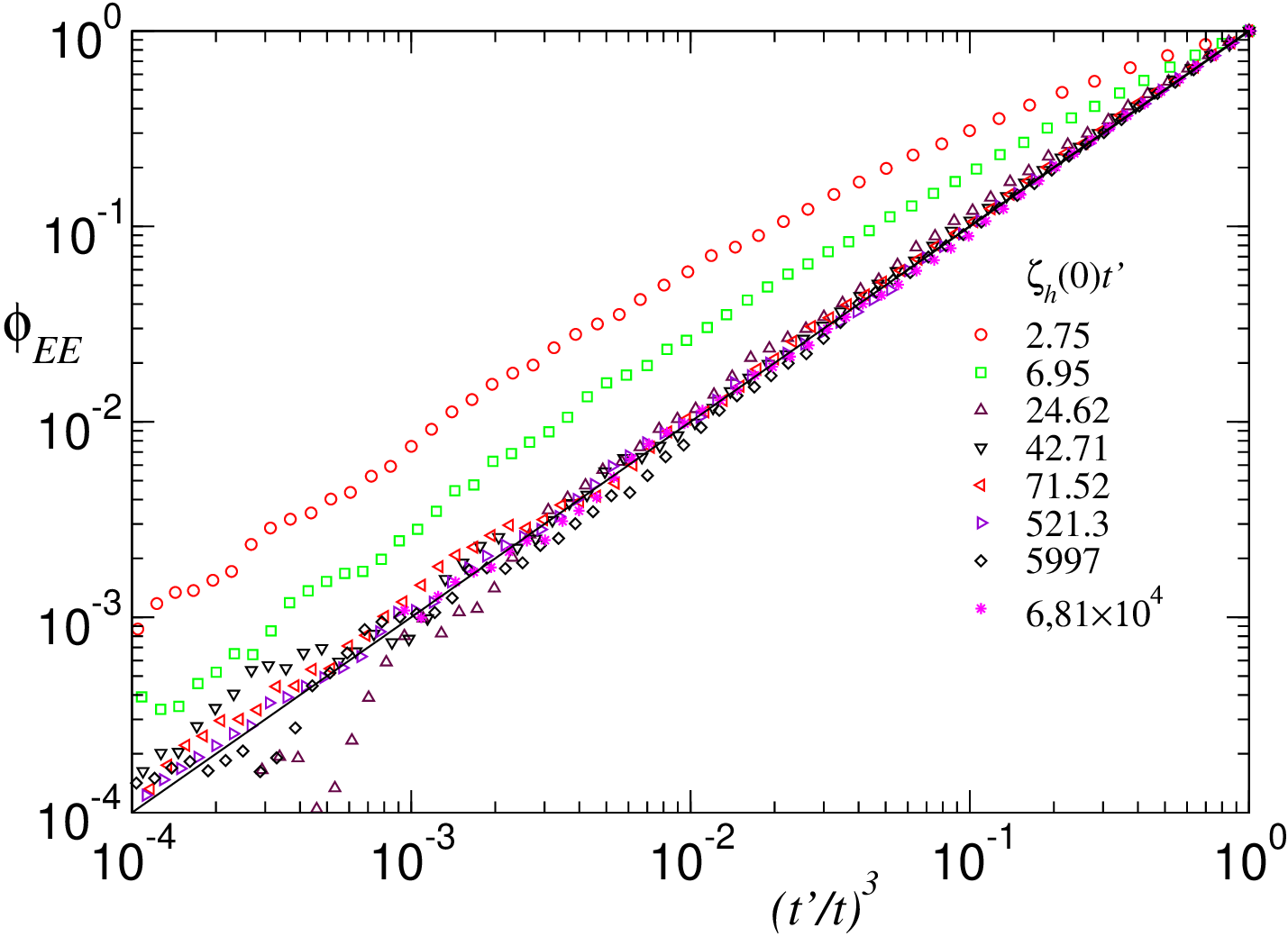}
\caption{Dimensionless time self-correlation function of the total
energy, $\phi_{EE}(t,t^{\prime})$, for a dilute granular gas of
inelastic hard disks in the HCS. The density is $n_{h}
\sigma^{2}=0.02$ and the restitution coefficient $\alpha=0.95$. The
different symbols correspond to different values of the initial
time, as indicated. The solid line is the theoretical prediction
describing the full aging phenomenon, Eq. (\ref{4.4}). \label{fig1}}
\end{figure}

To check the above theoretical predictions, we have performed MD
simulations of a system of inelastic hard disks ($d=2$). In Fig.
\ref{fig1}, results obtained for a system of $N=10^{3}$ particles
with  coefficient of normal restitution $\alpha=0.95$ and density
$n_{h} \sigma^{2}=0.02$ are reported. For these values of the
parameters, the HCS is stable, since the critical length is larger
than the size of the system and, therefore, the velocity vortices
and high density clusters characteristic of the clustering
instability \cite{GyZ93} cannot develop. Of course, in all the
simulations it has been verified that the system remains
homogeneous. The reported results have been averaged over 1200
trajectories of the system. The several curves correspond to
different values of $\zeta_{h}(0) t^{\prime}$, as indicated in the
figure. The value of $\zeta_{h}(0)$ has been estimated by using the
low density expressions derived in \cite{GyS95}.

In agreement with the analysis presented here, it is observed that
as $t^{\prime}$ increases the time-correlation function approaches a
form that depends only on the value of the time ratio
$t/t^{\prime}$. More precisely, for $\zeta_{h}(0) t^{\prime} \gtrsim
24$, all the plotted curves coincide within the statistical
uncertainties. Note that the latter increase as the value of $t$
grows, for a given value of $t^{\prime}$. Moreover, the asymptotic
curve agrees with the theoretical prediction given by Eq.
(\ref{4.4}), whose graphical representation is the solid line in the
figure. This good accuracy is consistent with the value of the
density of the system, that is small enough as to expect a low
density description, at the level of the Boltzmann equation, to
apply.

\begin{figure}
\includegraphics[scale=0.5]{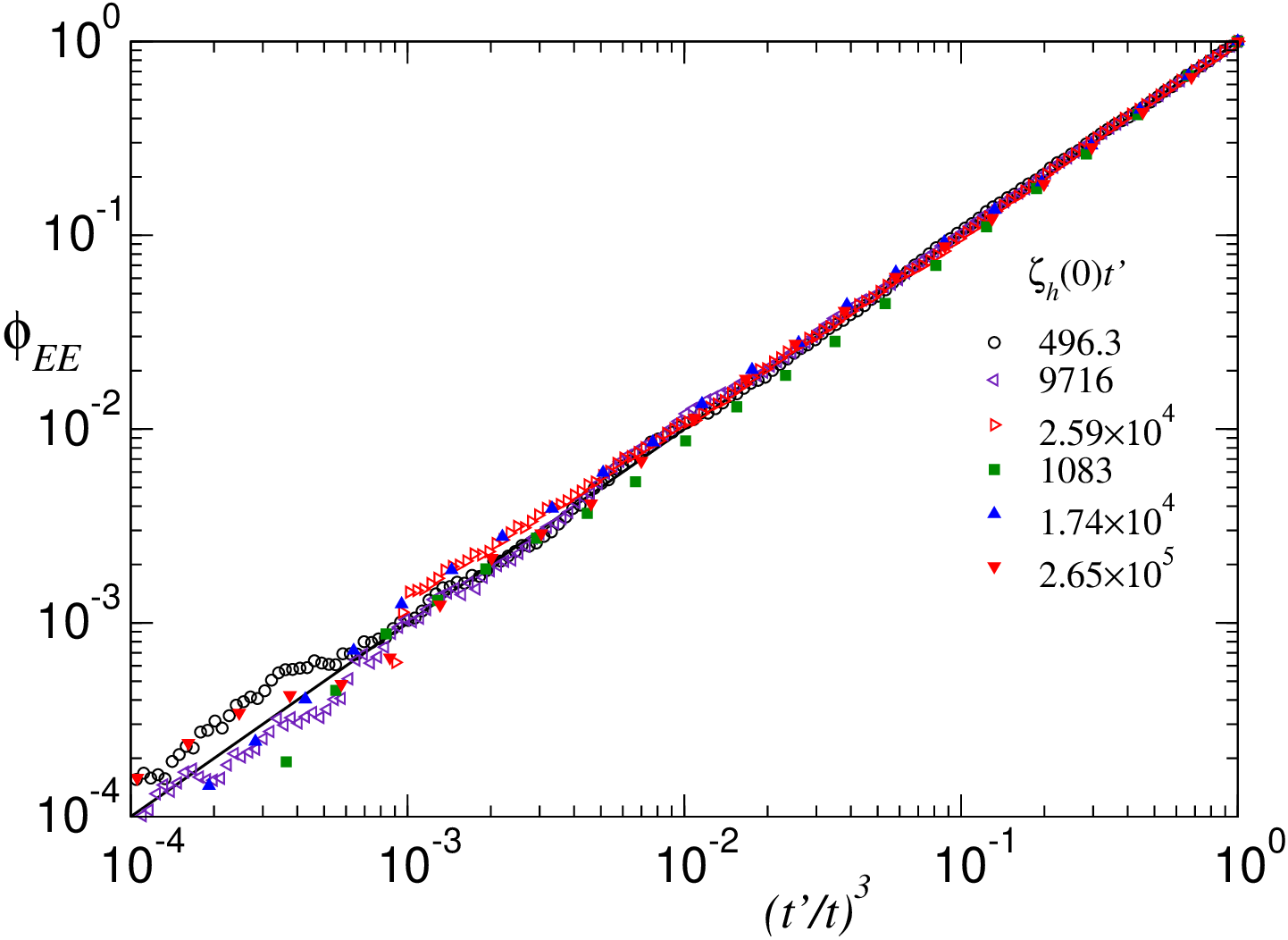}
\caption{The same as in Fig. \protect{\ref{fig1}}, but for different
values of the (low) density and of the restitution coefficient.
Filled symbols correspond to a system with $n_{h} \sigma^{2} =0.02$
and $\alpha = 0.85$, while empty symbols refer to a system with
$n_{h} \sigma^{2} = 0.1$ and $\alpha = 0.98$.  \label{fig2}}
\end{figure}

In Fig. \ref{fig2}, a similar plot is given, but now two systems,
one with $n_{h} \sigma^{2} = 0.02$, $\alpha = 0.85$, and the other
with $n_{h} \sigma^{2} =0.1$, $\alpha = 0.98$, are considered. For
the sake of clearness, only simulation data corresponding to large
waiting times have been included. Again, a good agreement with the
behavior predicted by Eq.\ (\ref{4.4}) is observed. The same
behavior has been obtained for other densities between the two above
values. This confirms the independence of $\phi_{EE}(t,t^{\prime})$
from the density and the inelasticity at low density. In fact, the
good agreement observed for $n_{h} \sigma^{2} =0.1$ must be
stressed, because at this density the gas can not be considered as
very dilute.

\begin{figure}
\includegraphics[scale=0.5]{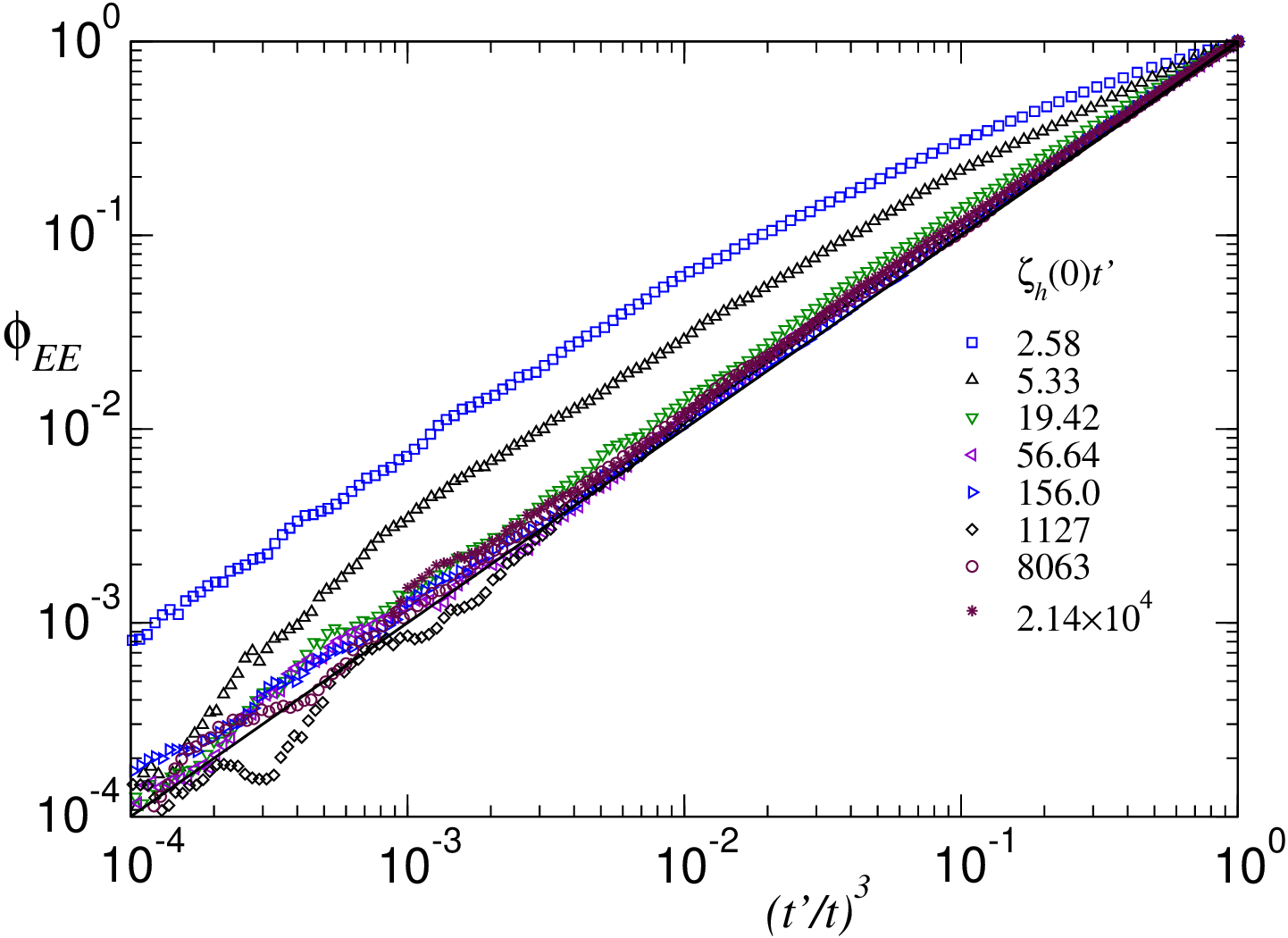}
\caption{The same as in Fig. \protect{\ref{fig1}}, but for $n_{h}
\sigma^{2}=0.2$ and $\alpha=0.98$. A clear tendency towards a
function depending only on $t/t^{\prime}$, characteristic of full
aging, is observed as $t^{\prime}$ increases. \label{fig3}}
\end{figure}

Prompted by the results for $n_{h}\sigma^{2} =0.1$ given in Fig.\
\ref{fig2}, the correlation function $\phi_{EE}(t,t^{\prime})$ has
also been evaluated at a definitely non small density, namely $n_{h}
\sigma^{2} =0.2$. For this value, density corrections to the low
density behavior are clearly identified in most of the equilibrium
properties of a molecular gas. It must be mentioned that, in order
to keep the system well inside the stable region of parameters and
with a number large enough of particles, the value of $\alpha$ must
be rather close to unity.  Moreover, they have been averaged over
1500 trajectories. The results shown in Fig. \ref{fig3} have been
obtained with $\alpha=0.98$ and $N=700$. Once again, a tendency
towards a behavior depending only on the ratio $t/t^{\prime}$ as
$t^{\prime}$ increases is clearly identified, i.e. the system
exhibits full aging. Besides, and rather surprisingly, the aging
phenomenon seems to be accurately described over several decades by
the law $\left( t^{\prime} / t \right)^{3}$, that was obtained here
in the context of very dilute granular gases (Boltzmann limit).

\section{Discussion}
\label{s5} The objective here has been to explore the existence of
aging in the homogeneous cooling state of a granular gas. This has
been done by exploiting the assumed scaling property of the
$N$-particle distribution of this state for a system of inelastic
hard spheres or disks. In fact, the presence of aging and its
specific form turn out to be directly associated to the scaling
property of the distribution \cite{DByL02,BGMyR04}. From this
perspective, the agreement between molecular dynamics simulations
and the theory discussed in this paper, is an almost direct proof of
the existence of the homogeneous cooling state also at the level of
the full many body pseudo-Liouville equation. This would extend the
rather well established fact that the inelastic Boltzmann and Enskog
equations have such a solution, consistently with some previous
results \cite{Lu01,BRyM04}.

Granular media are inherently non-equilibrium systems due to the
lack of energy conservation in the interactions between grains. They
present a very rich phenomenology which, sometimes, is similar to
that of normal, molecular systems. Moreover, it has been verified in
the last years that the methods of kinetic theory and
non-equilibrium statistical mechanics developed for normal fluids,
can be extended to granular fluids, yielding to results having an
analogous structure. There are also significant differences, but
they are well understood as consequences of the inelasticity. One
characteristic feature of granular systems is that quite often they
exhibit the phenomena in a much simpler context than molecular
systems. This refers to both, theoretical and experimental views. In
this paper, the simplicity of the HCS of a granular fluid has
allowed the identification of aging and the derivation of its
explicit form in the dilute limit, Eq.\ (\ref{4.4}), for the
correlation of the total energy of the system. The analytical
expression, that corresponds to the so-called full aging,  has been
shown to be in perfect agreement with MD simulation results. In
particular, the long time limit of the normalized time-correlation
function of the total energy is independent of the inelasticity and
the density. Quite interestingly, the simulations indicate that this
independence seems to extend to densities beyond the dilute limit.

To really appreciate the results presented here, it is important to
differentiate between both the existence of aging and the specific,
particular, form of the law governing it. Equation (\ref{3.16})
implies the existence of full aging in the system, i.e. the
normalized time-correlation function of the properties $A$ and $B$
is not a function of the time difference $t-t^{\prime}$ and depends
on the time ratio $t/t^{\prime}$. The only necessary condition to
derive this equation is the existence of the HCS itself, as
discussed above. On the other hand, identification of the function
$F$ in Eq.\ (\ref{3.16}) requires more detailed additional analysis,
which has been carried out only in the low density limit up to now.
The above leads to Eq. (\ref{4.4}) in the particular case of the
properties $A$ and $B$ being both the total (kinetic) energy of the
system. In this sense, the results presented in Fig. \ref{fig3}
strongly support the existence of the HCS  and the scaling property
of its $N$-particle distribution function at high densities,
independently of whether or not the convergence occurs towards the
power law given by Eq.\ (\ref{4.4}), as suggested by the simulation
results.

In real granular gases, the restitution coefficient is not constant,
but it depends on the impact relative velocity. Then, the
distribution function of the HCS does not scale in the form given by
Eq.\ (\ref{2.16}) and, therefore, the discussion in the present
paper does not apply in principle, although it can provide an
accurate approximation to the actual behavior of the system. To be
more precise, consider a given model of granular gas with
velocity-dependent coefficient of normal restitution. Now, the
problem being addressed has two energy scales. One is the total
energy per particle or, equivalently, the cooling temperature
$T_{h}(t)$. The other energy scale, $\epsilon$, is fixed by some
property of the specific collision model. Define a dimensionless
parameter
\begin{equation}
\label{5.1} \epsilon^{*} \equiv \frac{\epsilon}{m
v_{0}^{2}(T_{h})}\, .
\end{equation}
For hard spheres, $\epsilon=0$ and so $\epsilon^{*}=0$. It is in
this limit when the distribution function of the HCS has the scaling
property (\ref{2.16}) \cite{DByB07}. For $\epsilon^{*}>0$, the
scaling does not hold exactly, but it can be an appropriate
description as long as $\epsilon^{*} \ll 1$, i.e. the interaction be
sufficiently hard and/or the kinetic energy of the particles be
sufficiently large.

\section{Acknowledgements}

This research was partially supported by the Ministerio de
Educaci\'{o}n y Ciencia (Spain) through Grant No. BFM2005-01398.

\end{document}